\documentclass[aps,twocolumn,groupedaddress,showpacs]{revtex4}

\usepackage{epsf,latexsym}
\usepackage{times}

\newcommand{\bra}[1]{\langle #1 |}
\newcommand{\ket}[1]{| #1 \rangle}

\newcommand{\be}{\begin{equation}}
\newcommand{\ee}{\end{equation}}
\newcommand{\bae}{\begin{eqnarray*}}
\newcommand{\eae}{\end{eqnarray*}}

\def\CC{{\rm\kern.24em \vrule width.04em height1.46ex depth-.07ex
    \kern-.30em C}}
\def\P{{\rm I\kern-.25em P}}

\def\bbbc{{\mathchoice {\setbox0=\hbox{$\displaystyle\rm C$}\hbox{\hbox
to0pt{\kern0.4\wd0\vrule height0.9\ht0\hss}\box0}}
{\setbox0=\hbox{$\textstyle\rm C$}\hbox{\hbox
to0pt{\kern0.4\wd0\vrule height0.9\ht0\hss}\box0}}
{\setbox0=\hbox{$\scriptstyle\rm C$}\hbox{\hbox
to0pt{\kern0.4\wd0\vrule height0.9\ht0\hss}\box0}}
{\setbox0=\hbox{$\scriptscriptstyle\rm C$}\hbox{\hbox
to0pt{\kern0.4\wd0\vrule height0.9\ht0\hss}\box0}}}}

\def\bbbz{{\mathchoice {\hbox{$\sf\textstyle Z\kern-0.4em Z$}}
{\hbox{$\sf\textstyle Z\kern-0.4em Z$}}
{\hbox{$\sf\scriptstyle Z\kern-0.3em Z$}}
{\hbox{$\sf\scriptscriptstyle Z\kern-0.2em Z$}}}}

\newcommand{\putfigs}[3]{$$\leavevmode\hbox{\epsfxsize=#2 cm \epsfysize=#3 cm 
\epsffile{#1.eps}}$$}

\newcommand{\putfig}[2]{$$\leavevmode\hbox{\epsfxsize=#2 cm
   \epsffile{#1.eps}}$$}

\begin{document}
\title{Ground-State  Entanglement in Interacting Bosonic Graphs}
\author{Paolo Giorda$^{1,3}$  and Paolo Zanardi$^{1,2,3}$}

\affiliation{$^1$ Institute for Scientific Interchange (ISI), Villa Gualino, Viale
Settimio Severo 65, I-10133 Torino, Italy}

\affiliation{$^2$ Department of Mechanical Engineering,
Massachusetts Institute of Technology, Cambridge Massachusetts 02139
}

\affiliation{Istituto Nazionale per la Fisica della Materia (INFM), UdR Torino-
Politecnico, 10129 Torino, Italy}

\begin{abstract}
We consider a collection of  bosonic modes  corresponding to the  vertices,
of a graph $\Gamma.$ Quantum tunneling can occur only  along
the edges  of $\Gamma$  and a local self-interaction term is present.
Quantum entanglement of one vertex with respect the rest of the graph is analyzed
in the ground-state of the system as a function of the tunneling amplitude $\tau.$ 
The  topology of $\Gamma$ plays a major  role in determining the tunneling amplitude $\tau^*$ which leads 
to the maximum  ground-state entanglement.
Whereas in most of the cases one finds the intuitively expected result $\tau^*=\infty$
we show that it there exists  a family of graphs for which the optimal value of $\tau$ is pushed down to a 
finite value. We also show that, for complete garphs,  our bi-partite entanglement provides useful insights in the analysis
of the cross-over between insulating and superfluid ground states
\end{abstract}

\maketitle

{\em Introduction}-- Entanglement measures quantify the strength of purely quantum correlations
between subsystems of a compound quantum system \cite{ent}.
In the last few years efforts in the new field of  quantum information science unveiled
how such correlations can be exploited as a  genuine resource for carrying out computational and
communication tasks beyond the reach of any classicaly operating device \cite{qip}.
More recently  several  studies pointed  out that the notion of quantum entanglement
can be a useful conceptual tool to investigate the complex properties of  quantum many-body systems;
in particular the role of entanglement has been analyzed in spin systems undergoing
 quantum i.e., ground-state phase transitions \cite{qpt}

In this paper we shall study a related problem: the entanglement behaviour
in the ground-state of  a system made of a {\em finite} number of bosonic modes bi-linearly coupled each-other
and with a repulsive self-interaction.
Such a system can by described by a {\em graph} \cite{graph} whose vertices are
associated with the bosonic modes themselves and whose edges correspond 
to the bilinear couplings i.e., tunneling. The corresponding quantum Hamiltonian is a Bose-Hubbard one \cite{bh}
that in the general case represents a very difficult many-body problem.
In order to effectively tackle this problem we will mostly focus on graphs with a small number of vertices 
i.e., four.  In spite of this simplification our analysis reveals a variety of features
that are expected  to be of  general validity.

In Ref \cite{zan_graph} it has been shown that, in absence of of non-linear self-interaction
ground-state  entanglement of one vertex with respect the rest of the lattice
contains information about the graph topology.
Moreover the graph topology has a subtle interplay with self-interaction in affecting the entangling
power  \cite{ep} of tunneling coupling in small graphs \cite{pgpz}.

In the following  we will adhere to the approach to  quantum entanglement
in systems of indistinguishable particles discussed in Refs \cite{ind_ent,gitt, enk, shi}.
A  complementary one is pursued in Refs. \cite{schli,pask,Li} and more recently in \cite{wise}.
In our  approach  the subsystems
are provided by bosonic {\em modes} and {\em not } by the particles; in particular even the notion of 
single particle entanglement makes sense.
This  mode-entanglement concept is an inherently second-quantized one: the tensor product structure
of the state-space necessary in order to define entanglement \cite{tps} is provided
by   identification of the bosonic (fermionic) Fock space with a set of linear oscillators
(qubits) associated to single-particle state vectors.

As it will be illustrated in the last part of the paper the bi-partite mode entanglement
we study is closely connected to local i.e., on-site, particle number variance.
Quite recently this quantity has been showed to
be a truy physical resource to overcome limitations imposed by mass superselection
constraints by teleportation \cite{schuch}. Another quantum-information theoretic
motivation to our work is to study how interaction and graph
topology affect the capability  of entanglement generation  i.e., the entangling power \cite{ep},\cite{ham},
by adiabatically turning on the tunneling parameter at zero-temperature 

{\em Preliminaries}--
By an   {\em interacting bosonic graph} we mean a collection of  bosonic modes
associated to the vertices of a graph $\Gamma:=(V, E)$ \cite{graph} whose dynamics is governed  by 
the following Bose-Hubbard hamiltonian \cite{bh}
\begin{equation}
H=- \tau\sum_{(i,j)\in E} (b_i^\dagger b_j+{\rm{h.c.}}) +\varepsilon \sum_{i\in V} n_i^2
\label{BH}
 \end{equation}
where $n_i:=b_i^\dagger b_i$ is the occupation number operator of  the $i$-th vertex of $\Gamma.$
Eq. (\ref{BH}) can describe an inomogenehous spatial structure in which bosonic particles e.g., ultra-cold atoms,
reside over a collection of sites  --the set $V$ of vertices of $\Gamma$-- and can tunnel among different
locations, with amplitude $\tau,$  across the edges $E$ of the graph. The non-linear part in Eq (\ref{BH}), wheighted by 
the interaction parameter $\varepsilon,$ accounts for the self-interaction of the modes when more than one particle
is present in the same site.

It is important to stress that the Hamiltonian (\ref{BH}) may describe  a variety of quantum physical systems:
photonic modes coupled by beam-splitters and with Kerr non-linearities, arrays of Josephson junctions, ultra-cold atoms
in some (inomogenehous) optical lattice \cite{bec_lattice}. 
For the sake of concreteness we will mostly use a language in which the bosonic modes 
are thought to be spatially localized e.g., optical-lattice sites.
These kind of systems have  been already considered in the quantum information literature
in Refs. \cite{bec_lattice,milb_twowells,chen, radu_pz,duan_ent, simon,hines,micheli}.

In ref. \cite{zan_graph} the ground-state  entanglement properties  have been analyzed for  
pure tunneling. The main purpose of this work 
is to extend those results to $\tau<\infty.$
In this regime the dynamics described by (\ref{BH}) is a complex one due to the presence of two competing effects.
The tunneling term in (\ref{BH}) 
is responsible for  delocalizing the particles over the graph vertices, whereas the non-linear
coupling term tends to localize them. This interplay is dramatically displayed by the occurence of the 
superfluid-insulator transition predicted by the Hamiltonian (\ref{BH}) over lattices with a large number 
of sites \cite{bh} for a critical value of the coupling strength $\varepsilon.$  
One is then naturally led to conjecture that the higher the tunneling amplitude
$\tau$ the greater a single mode gets entangled with the rest of the graph.
We will see that that turns out to be the case for many of the possible graph topologies
but that it also exist topologies for which an increasing of $\tau$ results
in an {\em decreasing} of the mode entanglement of one vertex.

{\em Mode-entanglement and self-interaction}--
In this section we give a qualitative description of the results of the computations for the different systems
under study.
Our goal is to characterize, for the reference mode $0$, the behaviour of the mode entanglement 
of the ground state $|\Psi_{GS}(\tau)\rangle$ of Hamiltonians of the kind (\ref{BH})
when the ratio between the hopping parameter and the self-interaction parameter 
$\tau / \varepsilon$ varies from zero to a value that is much greater then one. 
In order to do so we fix $\varepsilon = 1$, that is we measure $\tau$ in  $\varepsilon$ units 
and let $\tau$ vary in $[0,\tau_{max}]$ ($\tau_{max} > 0$) with step 
$\triangle \tau$. For each value of $\tau$ we compute $E(\tau)={\rm{Tr}}[\rho_0(\tau) \log \rho_0(\tau)]/log(N+1)$,
where $\rho_0(\tau)=\rm{Tr}_{V_{\backslash 0}}(\ket{\Psi(\tau)}\bra{\Psi(\tau)})$ i.e., the reduced density matrix of the mode
$0$ \cite{pgpz}, the logarithms are taken in base two and the factor $1/\log(N+1)$ gives the normalization
of the mode entanglement to its maximum possible value.
In our first simulations we have considered rooted graphs $\Gamma_j$ 
with $N=L=4$, see figure ($1$); their Hamiltonian is given by (\ref{BH}) where the tunneling is allowed 
only between the sites linked by the edges of the relative graph.

We first focus on  the interval in which $\tau \ll 1$.
An intersting feature that can be highlighted in this regime, in which the tunneling part of the Hamiltonian can
be considered a perturbation of the self-interaction part, is the ordering of the curves with respect to the 
graph topology. This feature is particularly clear in fig. $(4)$ where the graphs $\Gamma_{j}$, $j=11,12,13$
can be obtained by adding respectively $1,2 \mbox{ and }3$ links to the sub-graph of $\Gamma_{10}$.
One can see that the greater the connectivity of the sub-graph $\Gamma_j-\{0\}$
the greater the entanglement, that is $E_{\Gamma_{13}}>E_{\Gamma_{12}}>E_{\Gamma_{11}}>E_{\Gamma_{10}}.$
The same ordering appears for the other two sets of graphs.
One finds $E_{\Gamma_{5}}>E_{\Gamma_{3}}>E_{\Gamma_{4}},$
where the sub-graphs of $\Gamma_3$ and $\Gamma_4$ have both two links; while, see figure ($3$),
$E_{\Gamma_{9}}>E_{\Gamma_{7}}>E_{\Gamma_{8}}>E_{\Gamma_{6}},$
and in this case the sub-graphs of $\Gamma_7$ and $\Gamma_8$ have the same number of links i.e, two.
The ordering of the graphs according their ground-state entanglement for small tunneling coupling $\tau$
can be related to the spectrum of the one-particle tunneling Hamiltoanian i.e, the spectrum of the adiacency matrix
of the graph. Indeed by diagonalizing the adiacency matrix of all the $L=4$ graphs we found that the greater
(in modulus) the minimum eigenvalue $\varepsilon_0$ the greater the associated entanglement.
This seems to be a natural and general result. Indeed $|\varepsilon_0|$  gives the strength of the tunneling rate
in the ground-state with no self-interaction, once $\tau$ is turned on, one expectes the full ground-state
to be mostly a coherent mixing of the  the ground state with $\tau=0$ i.e., the   state with one particle per vertex,
and the pure tunneling ground-state. The greater $\varepsilon_0$ the greater the wheight of this latter.

Noticeably in the regime $\tau>>1$ {\em the ordering of the curves is inverted}. This feature is again very clear for the 
set of graphs $\Gamma_j$, $j=10,11,12,13$; in fact, as we can see in fig. ($4$),
$E_{\Gamma_{10}}>E_{\Gamma_{11}}>E_{\Gamma_{12}}>E_{\Gamma_{13}}$
i.e., the greater the connectivity
the lower the mode entanglement of the mode $0$. This behaviour starts to be evident when 
$\tau \ge 1$ and is mantained even in the
asymptotic regime ($\tau \gg 1$) where now it is the self-interacion part of the Hamiltonian 
that plays the role of the perturbation. The latter feature can be seen by looking at the 
inset of figure ($4$) where it is displayed the behaviour of $E(\tau)$ over the full interval
$[0,\tau_{max}]$, $\tau_{max}=20$; when $\tau \gg 1$ the ordering of the curves remains 
the same described for $\tau \ge 1$.
For the set of graphs described in figure ($3$) we have the same kind of behaviour: the graph
with the lower connectivity $\Gamma_6$(one link in the sub-graph) presents the highest value of
$E(\tau)$, while $\Gamma_9$, the graph with the higher connectivity (three links in the sub-graph)
displays the lowest values of $E(\tau)$. In this case of $\Gamma_7$ and $\Gamma_8$ the ordering in
the asymptotic regime remains the same displayed for $\tau \ll 1$ i.e., $E_{\Gamma_{7}}>E_{\Gamma_{8}}$.

We treat separately the results (see figure $2$) for the graphs of the set $a)$ because in this 
case an interesting  behaviour comes into play. Here the ordering of the curves for $\tau \gg 1$ is of the same
kind of the one seen for the other set of graphs: $\Gamma_5$, the graph with the higher 
connectivity (three links in the sub-graph) has the lowest values of $E(\tau)$. But the very interesting
feature is that, {\em whereas  for all the other graphs  we have seen so far  $E(\tau)$ grows monotonically   as a function of 
$\tau$, in the case of $\Gamma_4$ and $\Gamma_5$ this is no longer true}. In fact what happens is that 
when $\tau$ starts to be different from zero $E(\tau)$ grows in both cases but then for higher values of $\tau$ 
it becomes a  decreasing function and reaches a stationary value for $\tau \gg 1$.
This behaviour is more evident for $\Gamma_5$, for which $E(\tau)$ presents a maximum for $\tau = 1.14$,
but it is also characteristic of $\Gamma_4$, for which $E(\tau)$ presents a maximum for $\tau = 3.22$.
\begin{figure}
\putfigs{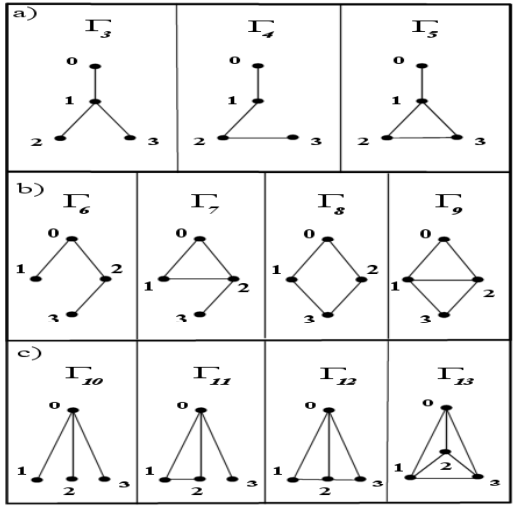}{7}{7}
\caption{Inequivalent rooted graphs for $N=L=4$; the root vertex $0$ is directly
linked with 1 (a), 2 (b) and 3 (c) vertices.}
\end{figure}
\begin{figure}
\putfig{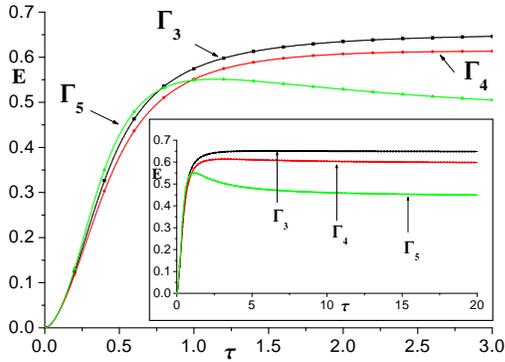}{7}
\caption{Plot of $E(\tau)$ for the systems $\Gamma_{j}$,$j=3,4,5$.}
\end{figure}
\begin{figure}
\putfig{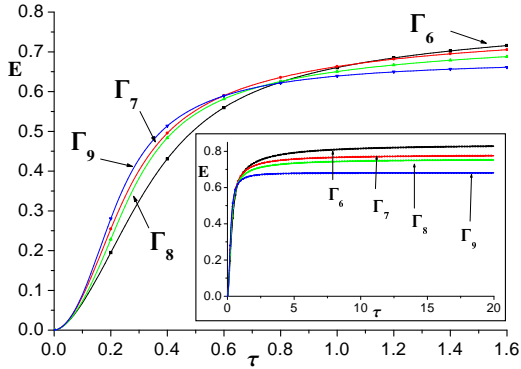}{7}
\caption{Plot of $E(\tau)$ for the systems $\Gamma_{j}$,$j=6,7,8,9$.}
\end{figure}
\begin{figure}
\putfig{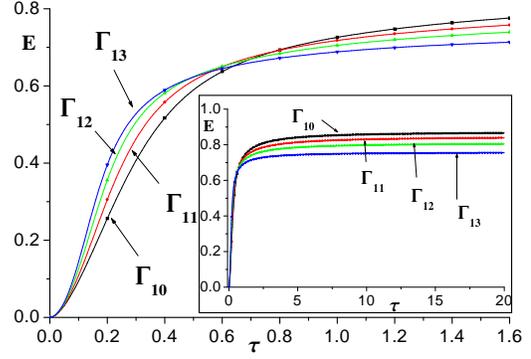}{7}
\caption{Plot of $E(\tau)$ for the systems $\Gamma_{j}$,$j=10,11,12,13$.}
\end{figure}
This peculiar behaviour can be somehow understood analytically in the following way. Let us  consider for example
the system represented by $\Gamma_5$; the latter belongs to a class of systems whose Hamiltonian, 
in the pure tunneling regime ($\varepsilon =0$), can be written as
$H= -\tau(b_0^\dagger b_1+{\rm{h.c.}}) -\tau\sum_{i\neq j=1}^{L-1} b_i^\dagger b_j.
$
It  corresponds to the topology $\Gamma$ in which the vertex $0$ is connected only with  
the vertex $1$ whereas the subgraph with vertices $\{1,\ldots,L\}$ is a complete one.
Now we provide a simple  argument showing that, in the large tunneling amplitude
limit and for a large number of vertices, 
the ground-state entanglement of the vertex $0$ for the systems described by the
above Hamiltonian is vanishing.

By introducing the Fourier operators ${\tilde b}_k:=1/\sqrt{L-1}\sum_{j=1}^{L-1} \exp( \frac{2\pi k j}{L-1}) b_j\;
(k=0,\ldots,L-2)$
the Hamiltonian associated to $\Gamma_5$ (with $\varepsilon=0)$ can be rewritten as $H =- \tau(L-1)(H_0+H_1),$ where
$H_0 = {\tilde b}_0^\dagger{\tilde b}_0 ,\; H_1=  (L-1)^{-1} [(b_0^\dagger b_1+{\rm{h.c.}})-N_{L-1}],
$
where $N_{L-1}:= \sum_{j=1}^{L-1} b_j^\dagger b_j.$
 The second term in the equation above can be regarded, for $L\mapsto\infty,$ as a small perturbation of the first
with coupling constant$\sim L^{-1}.$
It follows that the ground state of $H_0$  with $N=L$ particles, for $L=\infty$ is given by a condensate over the 
mode ${\tilde b}_0,$ i.e., $|\Psi_{GS}\rangle_{L=\infty} \sim  ({\tilde b}_0^\dagger)^L|0\rangle.$ In this latter state the 
mode $0$ clearly factors out with zero occupation number, therefore we have vanishing mode entanglement.
A simple first-order pertubation evaluation  gives 
$|\Psi_{GS}\rangle_L  \sim |\Psi_{GS}\rangle_{L=\infty} - L^{-1} b_0^\dagger {\tilde b}_0^{\dagger(L-1)}|0\rangle$  
which shows that, for large enough $L,$ the entanglement of the zero mode is a monotonically decreasing function of $L.$
This large $L$ behaviour is anticipated by the system $\Gamma_5$: here $L$ is finite and small, therefore $E(\tau)$
does not go to zero but it decreases and it reaches a finite non-zero value in the asymptotic regime i.e.,
$\tau \gg 1.$ The result presented above is robust against the turning on of a 
small self-interaction coupling $\varepsilon\neq 0$
Indeed even such a terms would be order $L^{-1}$ with respect $H_0$ \cite{gen}

This discussion indicates   that the size of  lowest single particle eigenvalue of $\Gamma-\{0\}$ 
plays a major role in determining the  entanglement properties we  analyze in this paper. This quantity in turn is well-known to have a clear 
topological meaning, for example for a regular graph with order $r$ vertices one has 
$\varepsilon_0= -r$ \cite{graph}. Roughly speaking, the greater the connectivity of $\Gamma-\{0\}$
the greater $|\varepsilon_0|.$

\begin{figure}
\putfig{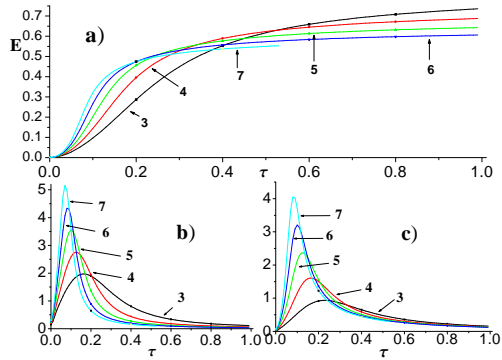}{7}
\caption{Plot of :  a) $E(\tau)$,  b) derivative of $E(\tau)$, c) derivative of 
$\langle n_i^2\rangle-\langle n_i\rangle^2$ for the complete graphs with $j=3,4,5,6,7$ }
\end{figure}

{\em Insultator-Superfluid cross-over.}--
In this section we 
show that the kind of bi-partite entanglement we analyzed in this paper provides useful insights
on the itinerant vs localized character of the particles in the ground state.
In particular  $E(\tau)$ can be related to the local particle number variance that it is
a standard tool to study the insulator-superfluid transition \cite{bec_mott}. 
We give a description of the simulation results of for the  case of   complete graphs.
To start with one can consider the simplest possible case, given  by $L=2$ i.e., the bosonic dimer \cite{hines}.
It is elementary to see that the ground state of (\ref{BH}) with $\varepsilon=1,$ given then given by 
$|\Psi_{GS}(\tau)\rangle:=\cos(\theta/2)|11\rangle +\sin(\theta/2)(|02\rangle+|20\rangle)/\sqrt{2},$
where $\theta=-\tan^{-1}(2\tau/\varepsilon)$
from which it follows immediately that the ground-state entanglement is given by
$E(\tau)= -\cos^2(\theta/2) \ln \cos^2(\theta/2) -\sin^2(\theta/2)\ln(\sin^2(\theta/2)/2).$
It is imeediate to see that $E(\tau)$  is a monotonic increasing function of $\tau;$
maximal entanglement is then achieved in the pure tunneling regime.

In order to measure the itinerant character of the particles over the graph
is useful to analyze the ground-state variance of the local occupation numbers
i.e., $\langle n_i^2\rangle-\langle n_i\rangle^2, (i\in V).$
This quantity plays the role of a sort of order parameter in the insulator-superfluid 
transition: small (large) values of it are associated to an insulator (superfluid).  
In the dimer case one gets the variance 
$\sin^2(\theta)=1/2(1-1/\sqrt{1+4\tau^2})$ that it is again a  monotonic increasing function of $\tau$;
its derivative shows a  peak for $\tau=1/2\sqrt{2}.$
The same kind of qualitative behaviour is displayed by $E(\tau)$ and by its derivative.

We considered complete graphs corresponding an increasing number of sites to i.e., $N=L=3,4,5,6,7$, see figure ($5$). 
In this case $E(\tau)$, which is plotted in the graphics $a)$, increses monotonically 
for all the interval $[0,\tau_{max}]$, $\tau_{max}=1$. 
In the inset of the figure ($5$) it is plotted the variance $\langle n_i^2\rangle-\langle n_i\rangle^2, (i\in V)$,
that, in view of the symmetry of the systems, is the same for all the sites $i$. As for the mode entanglement
it increases monotonically for all the interval $[0,\tau_{max}]$.
We have therefore plotted the derivative of this quantity  and we have compared it with the derivative of the mode entanglement, 
see graphics $c)$ and $b)$ in figure ($5$). Of course the insulator-superfluid  phase transition occurs only in the thermodynamic limit 
(when $N,L \rightarrow \infty$) so the peak in the derivative plot can only be regarded  as a far precursor of this transition. 
In this case, the derivative exibits the behaviour expected by such a precursor, in fact, as $N=L$ increases it becomes
more and more peaked; in presence of an actual phase transition it should display a singular behaviour 
in correspondance of a critical value $\tau_c$ of the tunneling parameter. 
These results clearly show that $E(\tau)$ has the same qualitative behaviour of the variance the local
particle variance \cite{comp}  and therefore, at least for complete graphs,  {\em it represents on itself a useful quantity to analyze 
the cross-over between insulating and super-fluid phases}. We observe that this result is reminescent
of  an analog one for the metal-insulator transition in the {\em Fermionic} Hubbard nodel \cite{you1}. 

{\em Conclusions.}--
We have  analyzed ground-state entanglement of interacting bosons for all the four vertices  graphs.
The ground-states for all possible graph topologies  have been obtained
by numerical diagonalization and the entanglement of one vertex with respect to all the others
has been computed as a function of the  tunneling amplitude.
This analysis allows to order graphs, for a given tunneling amplitude $\tau,$ in terms
of the amount of this bi-partite ground-state entanglement $E(\tau).$
Remarkably this  order actually depends on both $\tau$ and the graph topology: 
for small (large) $\tau$ the higher (lower) the modulus of the minimum single-particle eigenvalue of the graph $\Gamma$
(sub-graph $\Gamma-\{0\}$)  
the higher (lower) the mode entanglement.
Moreover numerical results show un unexpected, non monotonic,
behaviour of $E(\tau)$ for some particular graph topologies. 
Such a phenomenon can be understood analytically in the limit of no self-interaction.
We finally showed, for complete graphs of different sizes, 
that $E(\tau)$ contains useful information about the cross-over
between the insulator and superfuild regime. 
The analysis of the role of topology and self-interaction on 
multi-partite entaglement, for example by considering all possible bi-partitions
of the graph at once, is a more demanding task clearly deserving future investigations 

P.Z. gratefully acknowledges financial support by Cambridge-MIT Institute Limited and by the 
European Union project  TOPQIP (Contract IST-2001-39215)


\begin{references}
\bibitem{gen} We can further generalize the above argument as follows. The Hamiltonian (\ref{BH}) can be written, for $\tau=\infty,$
as  a sum of two terms $H_0$ and $H_1;$ the first containing all the contributions involving the mode $0$ and the 
second describing just the sub-graph $\Gamma-\{0\}.$ Let $H_1=\sum_\alpha \varepsilon_\alpha {\tilde b}_\alpha^\dagger 
{\tilde b}_\alpha$
the diagonal form of $H_1.$ If we  define $\varepsilon_0 :=\min_\alpha  \varepsilon_\alpha$ as the
single-body ground state, the ground-state of $H_1,$ with $L$ particle is a a condensate over the corresponding
eigen-mode with zero particle in the $0$ vertex.
One  can write $H= \varepsilon_0( {\tilde b}_0^\dagger {\tilde b}_0+ 1/ \varepsilon_0 H_0)+ H^\prime,$
where $H^\prime$ is a sum over the excited single-body terms of $H_1.$ By denoting with $l_0$ the number
of edges linking  the vertex $0$
with the rest of the graph, the second term in the equation above  is at most of the order $l_0/\varepsilon_0.$
When the modulus of this  ratio gets very small e.g., $l_0=O(1)$ and $\varepsilon_0=O(L)$ for large $L,$
the $H_0$ contribution is a perturbative one and at zero temperature all the particles tend to be localized within the
sub-graph  $\Gamma-\{0\}$ i.e., no entanglement of the $0$ vertex with the others.
\bibitem{comp} This can be understood by observing that the reduced density matrix is a diagonal 
i.e., $\rho=\rm{diag}(\rho_i)_{i=0}^{N}$ and $E$ is just the (normalized) Shannon entropy of the
probability distribution $\rho_i;$ whereas the local particle number variance
is $\sum_i i^2 \rho_i -(\sum_i i \rho_i)^2$
\end{references}
\end{document}